\documentclass[twocolumn,tighten,twocolappendix]{aastex63}
\usepackage[T1]{fontenc}
\usepackage{graphicx}
\usepackage{amsmath}
\usepackage{amsfonts}
\usepackage{amssymb}
\usepackage{natbib}
\usepackage{textcomp}
\begin{document}

\title{Onset of turbulent fast magnetic reconnection observed in the solar atmosphere}

\author[0000-0002-9270-6785]{L. P. Chitta}
\affiliation{Max Planck Institute for Solar System Research, Justus-von-Liebig-Weg 3, D-37077 G\"ottingen, Germany; \href{mailto:chitta@mps.mpg.de}{chitta@mps.mpg.de}}

\author{A. Lazarian}
\affiliation{Astronomy Department, University of Wisconsin, Madison, WI 53706, USA; \href{mailto:alazarian@facstaff.wisc.edu}{alazarian@facstaff.wisc.edu}}

\begin{abstract}

Fast magnetic reconnection powers explosive events throughout the universe, from gamma-ray bursts to solar flares. Despite its importance, the onset of astrophysical fast reconnection is the subject of intense debate and remains an open question in plasma physics. Here we report high-cadence observations of two reconnection-driven solar microflares obtained by the \textit{Interface Region Imaging Spectrograph} that show persistent turbulent flows preceding flaring. The speeds of these flows are comparable to the local sound speed initially, suggesting the onset of fast reconnection in a highly turbulent plasma environment. Our results are in close quantitative agreement with the theory of turbulence-driven reconnection as well as with numerical simulations in which fast magnetic reconnection is induced by turbulence.

\end{abstract}

\keywords{Solar magnetic reconnection (1504); Solar atmosphere (1477); Solar magnetic fields (1503); Spectroscopy (1558); Plasma physics (2089); Magnetohydrodynamics (1964); Solar transition region (1532); Solar chromosphere (1479); Solar flares (1496); Solar active regions (1974); Solar ultraviolet emission (1533)}

\section{Introduction} \label{sec:intro}

Magnetic reconnection is a fundamental process throughout the universe \citep[][]{2000mare.book.....P} that converts magnetic energy to other forms of energy, driving solar flares \citep[][]{1966Natur.211..695S,1994Natur.371..495M,2013NatPh...9..489S}, gamma ray bursts \citep[][]{2019ApJ...882..184L}, and accelerating particles \citep[][]{2017PhRvL.118o5101K}. In spite of its importance, the nature of astrophysical fast reconnection, necessary to explain explosive flares and bursts, is the subject of intense debate. 2D numerical research has shown that the instabilities of a current layer to the formation of small magnetic islands, plasmoids, trigger fast reconnection \citep[e.g.][]{2007PhPl...14j0703L,2009PhPl...16k2102B,2017ApJ...849...75H}. At the same time, 3D numerical simulations suggest the development of turbulence in the reconnection layer \citep[][]{2019PhPl...26a2901D}. For reconnection layers much wider than the corresponding plasma scales, the magnetohydrodynamic (MHD) approximation is applicable. The 3D simulations in this regime \citep[][]{2009ApJ...700...63K,2017ApJ...838...91K,2019arXiv190909179K,2017ApJ...834...47B} suggest that the fast reconnection is induced by turbulence and the results correspond to turbulent reconnection theory \citep[][see also \citealt{lazarian2020} for a review]{1999ApJ...517..700L}, one of the dramatic predictions of which (the violation of the flux freezing in turbulent fluids) has been tested recently \citep[][]{2013Natur.497..466E}. Nevertheless, the key issue of the trigger of fast reconnection is not settled.
  
Fast magnetic reconnection is widespread in the solar atmosphere. It drives explosive events and ultraviolet (UV) bursts \citep[e.g.,][]{1991JGR....96.9399D,1997Natur.386..811I,2014Sci...346C.315P}, and small and large flares in general \citep[e.g.,][]{2010ApJ...712L.111J,2013NatPh...9..489S,2014ApJ...795L..24T,2015ApJ...812...11V,2018A&A...615L...9C}. Observations of solar bursts, jets and flares in (extreme) UV radiation show plasmoid-like blobs \citep[e.g.][]{2017ApJ...851L...6R,2019ApJ...885L..15K,2019ApJ...870..113Z}. However, due to the lack of magnetic field measurements in those observed blobs, their exact nature and direct connection to fast reconnection, particularly in naturally 3D systems, is unclear \citep[][]{2015SSRv..194..237L,2017ApJ...841...27N,2018SSRv..214..120Y}. On the other hand, turbulence is observed in pre-flare flux ropes that later flare \citep[][]{2013ApJ...774..122H}, and small- and large-scale current sheets \citep[e.g.,][see \citealt{2015SSRv..194..237L} for a review on current sheets]{2008ApJ...686.1372C,2018ApJ...866...64C,2018ApJ...854..122W,2018ApJ...858L...4X}. In addition, turbulence is proposed to be an energy transport process in solar flares \citep[e.g.][]{2017PhRvL.118o5101K,2018SciA....4.2794J}. Its persistence and characteristics directly at the reconnection site, and its role in triggering fast reconnection, however, remain unknown. 

Reconnection-driven microflares in the solar atmosphere typically occur in the cores of active regions, hosting magnetic field of high intensity \citep[][]{2010ApJ...712L.111J,2014ApJ...795L..24T,2015ApJ...812...11V}. Due to their compact size and short lifetimes compared to large-scale flares that follow a complex evolution over tens of minutes to hours, the microflaring loops are potential candidates to probe the physics of fast reconnection. We study plasma properties of two microflares to investigate the onset conditions of magnetic reconnection using  high-cadence UV spectroscopic and imaging observations from the \textit{Interface Region Imaging Spectrograph} \citep[\textit{IRIS};][]{2014SoPh..289.2733D}. 

\begin{figure*}[t!]
\begin{center}
\includegraphics[width=0.45\textwidth]{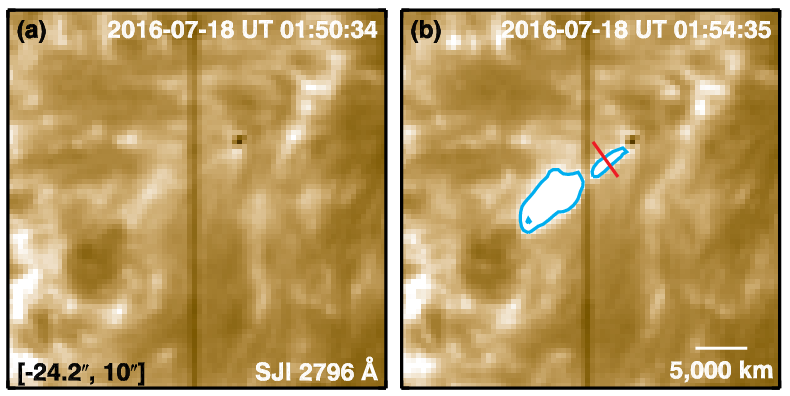}
\includegraphics[width=0.75\textwidth]{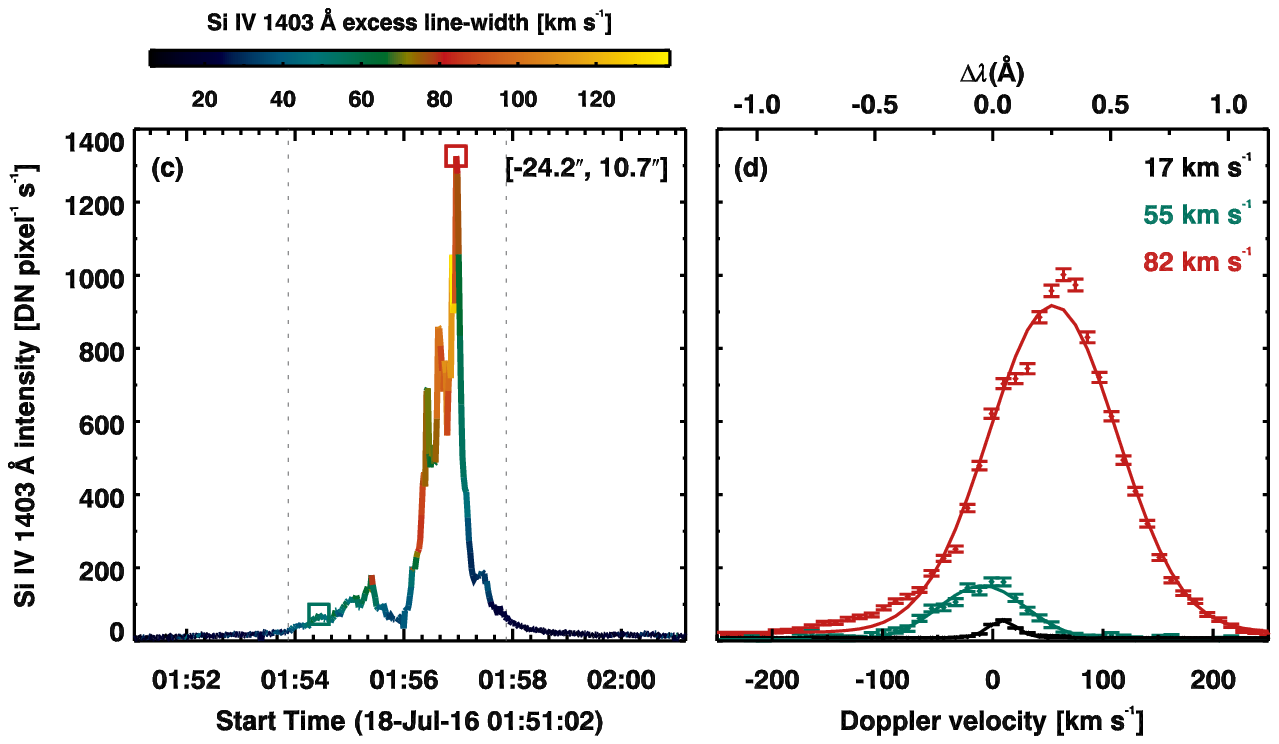}
\includegraphics[width=0.75\textwidth]{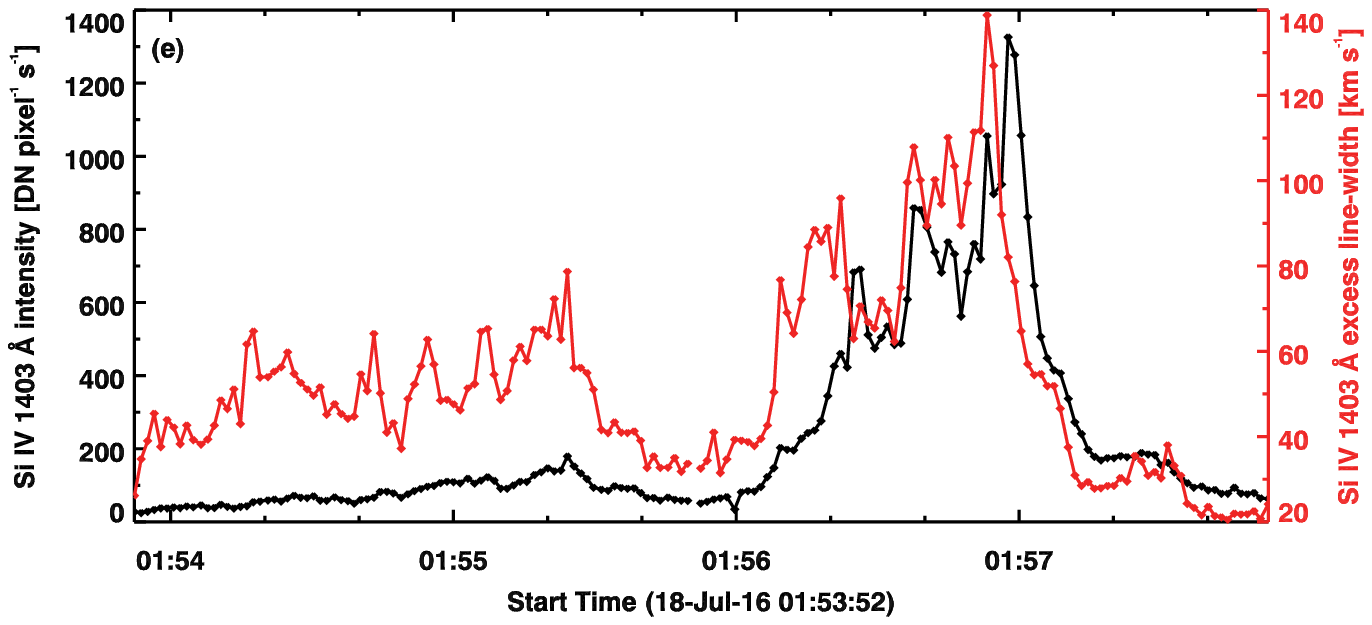}
\caption{Microflare in the core of AR 12567 observed on 2016 July 18. Panels (a) and (b): \textit{IRIS} SJI\,2796\,\AA\ snapshots sampling the chromosphere, centered at $(-24.2\arcsec, 10\arcsec)$, covering an area of $50\arcsec\times50\arcsec$ on the Sun. An animation of these panels is available. The video begins at 2016 July 18 01:50:34 UT and ends the same
day at 01:57:53 UT. The vertical line is the location of the \textit{IRIS} slit for spectroscopic observations. In panel (b), the slit crosses the center of the microflare. The slanted red line marking a cut across the loop segment is used for space-time map in Figure\,\ref{fig:space}. The cyan contour outlines the loop. Panel (c) shows plasma properties from the loop. The solid multi-colored curve is the time series of the Si iv 1403\,\AA\ line intensity. The curve is color-coded with the excess broadening of the Si\,{\sc iv} line. Panel (d) displays sample Si\,{\sc iv} spectral profiles at two instances of the brightening (marked by the similar colored squares in panel (c)). The symbols are the observed profiles and the corresponding photon noise. The solid curves are the single-Gaussian fits (red-colored profiles multiplied by 0.5). The black curve is the average Si\,{\sc iv} spectrum from a quiet region in that AR (multiplied by a factor of 3). Respective excess line widths of the line profile are listed. In the top axis, zero corresponds to the reference wavelength of the Si\,{\sc iv} line of 1402.77\,\AA. Panel (e) shows the intensity (black) and excess width of the Si iv 1403\,\AA\ line, separately as a function of time (between the two vertical dotted lines in panel (c) covering the microflare activity). See Section\,\ref{sec:obs} for details.\label{fig:cont1}}
\end{center}
\end{figure*}

\section{Observations} \label{sec:obs}

On 2016 July 18, \textit{IRIS} observed an active region (AR) 12567 located near disk center at $(-25\arcsec, 31\arcsec)$, in sit-and-stare mode, from 01:38\,UT to 02:38\,UT. During this period, \textit{IRIS} caught the rapid flaring of a 14\,Mm long loop, in the core of the AR, that lasted for about 4\,minutes. The brightening is imaged by the 2796\,\AA\ passband of \textit{IRIS} slit-jaw imager (SJI), that covers Mg\,{\sc ii} h and k lines (sampling chromospheric temperatures of about $10^4$\,K). The SJI images have a cadence of 14\,s. The spectrographic slit of \textit{IRIS} crossed the loop close to its center and captured the whole event at a high cadence of 1.4\,s, exposure time of 0.5\,s and a spatial sampling of 0.665\arcsec\,pixel$^{-1}$ along the slit (1\arcsec\ corresponds to about 725\,km on the Sun). We use the latest version of calibrated level-2 \textit{IRIS} data that include corrections for flat-fielding, dark currents, geometric distortions, scattered light and background, and wavelength calibration \citep[][]{2018SoPh..293..149W}. Here we focus on the spectroscopic observations of plasma emission from the Si\,{\sc iv}\ spectral line at 1403\,\AA\ (equilibrium formation temperature of 0.08\,MK), and the density-sensitive O\,{\sc iv} line pair at 1401 and 1405\,\AA\ (equilibrium formation temperature of 0.14\,MK). The spectral sampling is 50\,m\AA. Observational details of another microflare are discussed in Appendix\,\ref{sec:case2}. To study the surface magnetic field distribution underlying these microflares, we used data from the Helioseismic and Magnetic Imager \citep[HMI;][]{2012SoPh..275..207S} on board the \textit{Solar Dynamics Observatory} \citep[\textit{SDO};][]{2012SoPh..275....3P}. 

We characterize plasma properties of the loops using the Si\,{\sc iv} line parameters, in particular, its intensity and width. Intensity is derived by integrating the observed spectral line profile in the Doppler velocity range of $\pm250$\,km\,s$^{-1}$, corresponding to a wavelength range from 1401.6 to 1403.94\,\AA\, where the reference wavelength, $\lambda_0$, for Si\,{\sc iv} line is 1402.77\,\AA\ \citep[][]{1986ApJS...61..801S}. At each time-step the average continuum intensity in the window 1405.2 to 1406\,\AA\ is subtracted from the integrated Si\,{\sc iv} intensity. The integration is performed at each spatial pixel along the slit as a function of time.

We obtain the full-width at half maximum (FWHM), $\delta\lambda$, of the line by performing a single-Gaussian fit to the observed spectral profiles. Gaussian fitting is done through \texttt{eis\_auto\_fit.pro} available in IDL/solarsoft that employs MPFIT procedures \citep[][]{2009ASPC..411..251M}. The Gaussian FWHM has contribution from thermal broadening, $v_{\mathrm{th}}$, instrumental FWHM, $\sigma_\mathrm{I}$, and the residual broadening, $\xi$ (i.e., excess line width not accounted by thermal and instrumental broadenings). Thermal width is defined as $v_{\mathrm{th}}=\sqrt{2k_\mathrm{B}T_\mathrm{i}/m_\mathrm{i}}$, where $k_\mathrm{B}$ is Boltzmann's constant, $T_\mathrm{i}$ is the temperature of the ion, and $m_\mathrm{i}$ is its mass. The line width is then given by
\begin{equation}
\delta\lambda=\frac{\lambda_0}{c}\sqrt{4\mathrm{ln}\left(2\right)\left(v_{\mathrm{th}}^2+\xi^2\right)+\sigma_\mathrm{I}^2}.
\end{equation}
Here the nominal thermal width of the Si\,{\sc iv} line at equilibrium formation temperature of 0.08\,MK is about 6.6\,km\,s$^{-1}$. Under nonequilibrium conditions, the Si\,{\sc iv} line forms over a much broader range of temperatures between $10^4$ and $10^5$\,K, and outside this temperature range the line emissivity drops significantly \citep[][]{2016ApJ...817...46M}. The corresponding thermal speeds would then be in the range of 2.5\textendash7.5\,km\,s$^{-1}$, not significantly different from the thermal width of the Si\,{\sc iv} line under equilibrium conditions. In the rest of this Letter, we will use nominal thermal width of the Si\,{\sc iv} line to determine the excess line width. In the far-UV covering Si\,{\sc iv} line, the \textit{IRIS} instrumental width (FWHM) is about 0.026\,\AA, which corresponds to 5.5\,km\,s$^{-1}$ \citep[][]{2014SoPh..289.2733D}. By substituting the values for $\delta\lambda$ determined by Gaussian fitting, $v_{\mathrm{th}}$ and $\sigma_\mathrm{I}$, we obtained $\xi$ at each pixel along the slit as a function of time. Any values of $\xi$ due to observational artifact are discarded in our analysis. 

\section{Plasma and Magnetic Properties of the Microflaring Loop} \label{sec:prop}

\begin{figure}[ht!]
\begin{center}
\includegraphics[width=0.45\textwidth]{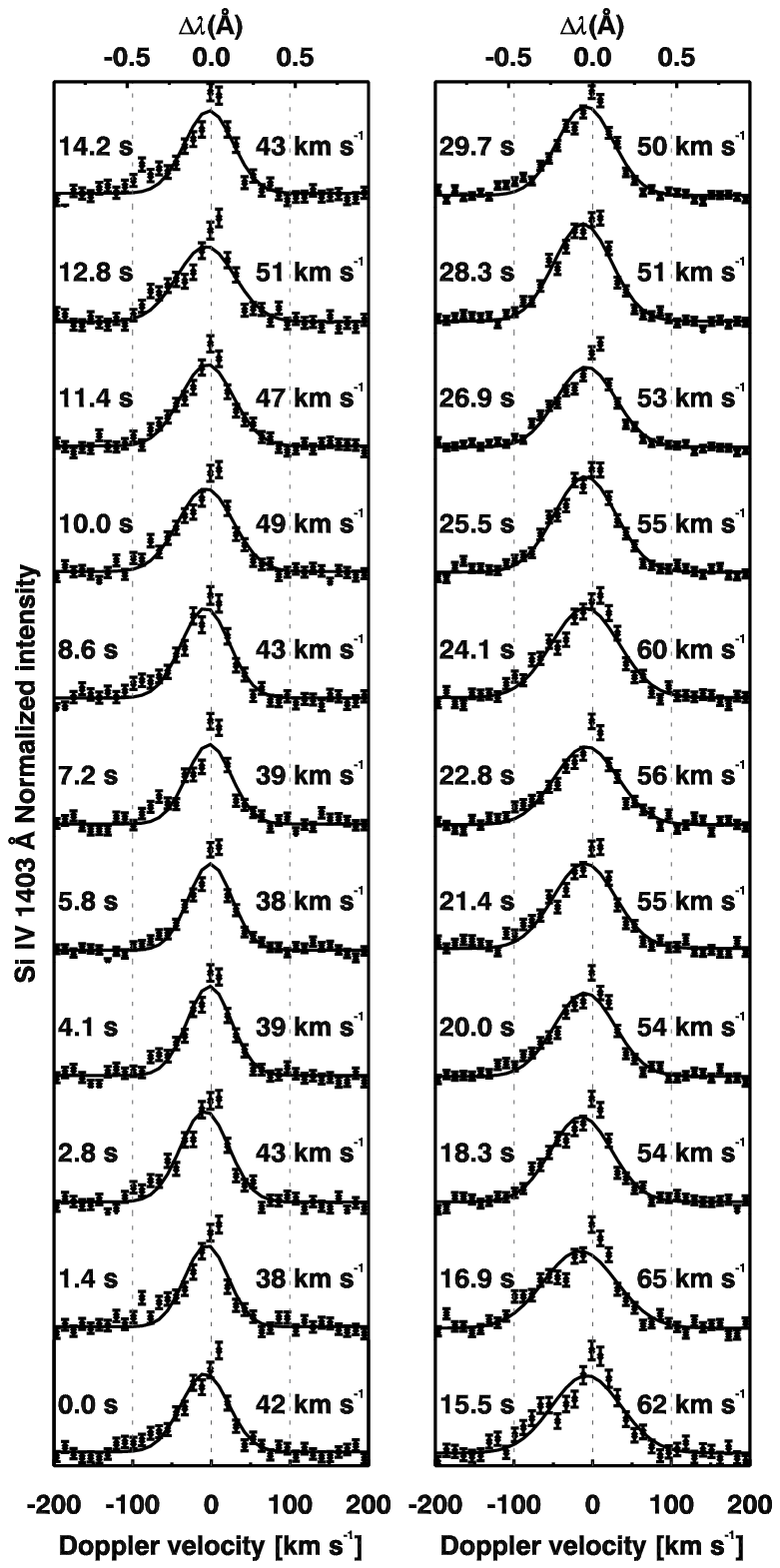}
\caption{Spectral evolution at the site of microflare. Each observed Si\,{\sc iv} spectral profile obtained at the site of loop brightening shown in Figure\,\ref{fig:cont1} is normalized to its peak and is plotted as a function of Doppler velocity. The symbols denote observations and the corresponding error bars are for the photon noise. The solid curves are the single-Gaussian fits to the observed profiles. The time in seconds elapsed since 01:54\,UT on 2016 July 18, and the excess line broadening are listed. Time increases from bottom to top and from left to right. The three vertical dotted lines mark the Doppler velocities of $-100$, $0$, and $100$\,km\,s$^{-1}$, with respect to the reference wavelength of Si\,{\sc iv} line.
The plot is continued in Figure\,\ref{fig:spect2} (see Sections\,\ref{sec:obs} and \ref{sec:prop} for details).\label{fig:spect1}}
\end{center}
\end{figure}

\begin{figure}[t!]
\begin{center}
\includegraphics[width=0.48\textwidth]{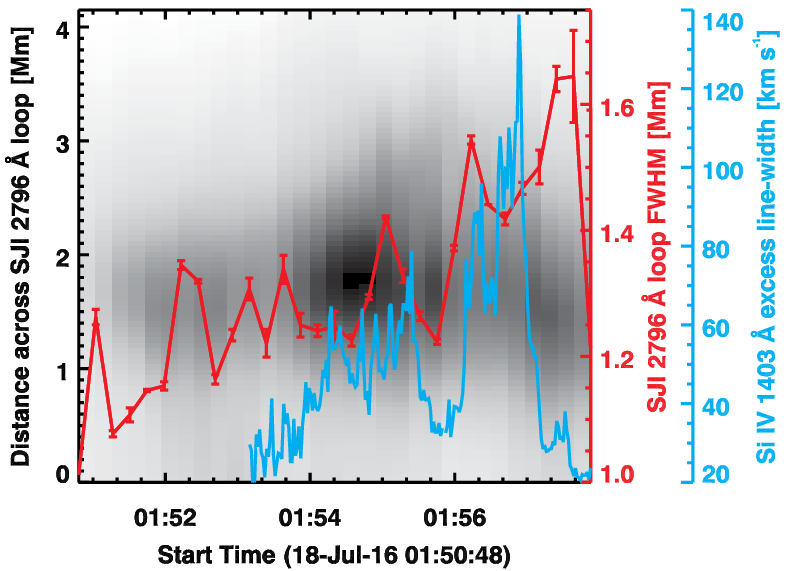}
\caption{Evolution of a segment of microflaring loop. The background negative image is obtained by stacking the \textit{IRIS} SJI intensity as a function of time along the cut (averaged over $\pm2\arcsec$ across the cut) shown in Figure\,\ref{fig:cont1}. The red curve is the FWHM of the loop observed in SJI data. The error bars are 1$\sigma$ standard deviation of the distribution of FWHMs obtained from a set of 1000 random realizations of photon noise. The cyan curve shows the evolution of excess line broadening of the Si\,{\sc iv} line profile at the location where the slit crosses the loop (see Figure\,\ref{fig:cont1}).\label{fig:space}}
\end{center}
\end{figure}

\begin{figure*}[t!]
\begin{center}
\includegraphics[width=0.4\textwidth]{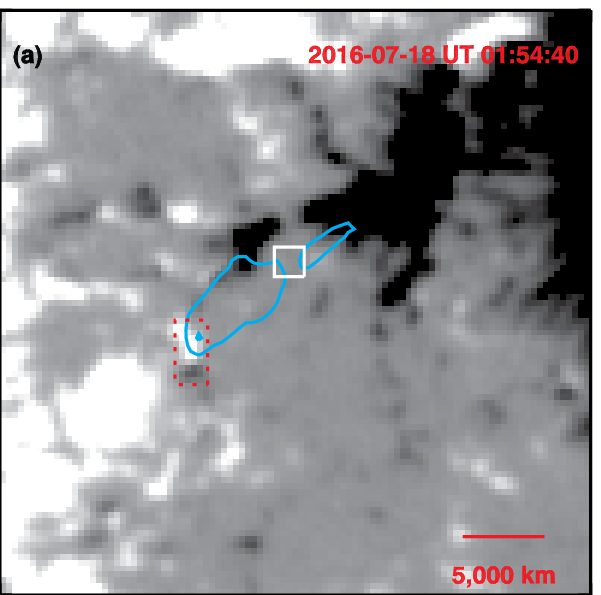}
\includegraphics[width=0.5\textwidth]{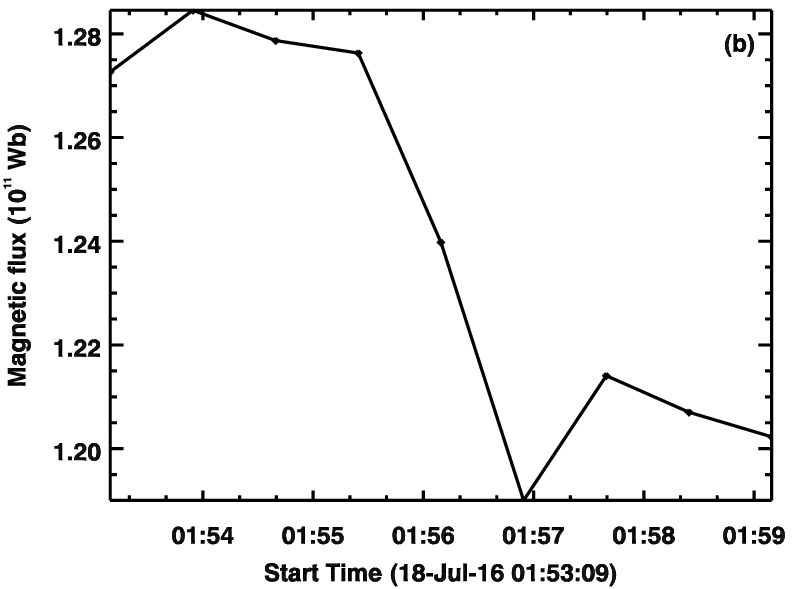}
\caption{Magnetic setting of the microflare. Panel (a): line-of-sight map of photospheric magnetic field obtained by \textit{SDO}/HMI in the core of AR 12567 is displayed. The field of view is the same as in Figure\,\ref{fig:cont1}. The light and dark regions represent positive and negative polarity magnetic fields (saturated at $\pm0.03$\,T). The cyan contour outlines the overlying loop (see Figure\,\ref{fig:cont1}). The white box intersects the center of the loop in the line of sight. The region enclosed by the box is used to calculate the height variation of magnetic field strength above the solar surface (see Figure\,\ref{fig:denmag}). The red dotted rectangle (length of 4\,Mm) at the eastern footpoint covers a magnetic bipole. Panel (b): Time series of integrated unsigned magnetic flux enclosed by the red rectangle in panel (a) at the eastern footpoint. To avoid noise, we considered only those pixels with magnetic flux density above $10^{-3}$\,T for integration.\label{fig:mag}}
\end{center}
\end{figure*}

A 14\,Mm long microflare is observed by \textit{IRIS} SJI with 2796\,\AA\ passband, in the core of active region AR\,12567 on 2016 July 18 (top panels of Figure\,\ref{fig:cont1}). The microflare feature is composed of a loop system rooted in multiple mixed magnetic polarities at the photosphere (see Figure\,\ref{fig:mag}). The integrated intensity of the Si\,{\sc iv}\ spectral line recorded by the \textit{IRIS} spectrograph at the center of the loop system shows more than two orders of magnitude fluctuation over the course of 4\,minutes (solid multi-colored curve in Figure\,\ref{fig:cont1}(c)). The intensity fluctuations of the microflare are structured in two phases. In the first, weaker and gradual phase that lasted for 2\,minutes, intensity gently increases and decreases. Its overall increase is an order of magnitude more than the background (between 01:54\,UT and 01:56\,UT). The second, stronger phase is more abrupt where the intensity increases from the background level to its peak (more than two orders of magnitude change) in 60\,s. This rapid rise phase is further superimposed with multiple, temporally resolved intensity fluctuations that lasted 5 to 10\,s each. After the event, the intensity decreases and falls back to the background level.

We characterize the plasma flows in the microflaring loop system using the width of the Si\,{\sc iv}\ emission line, determined by a single-Gaussian fit (see Section\,\ref{sec:obs}). The $1/e$ width of the Si\,{\sc iv}\ line in the loop is significantly larger than its nominal thermal width of 6.6\,km\,s$^{-1}$ (Figure\,\ref{fig:cont1}(d)). This excess width should arise from either unresolved non-thermal motions or superimposed bulk plasma flows with a broad distribution of line-of-sight velocities. In either case, these flows are mostly confined to the microflaring loop system itself and are directed perpendicular to its long (magnetic) axis. Thus, the broad nature of the observed emission here characterizes turbulent flow system internal to the loop. These turbulent flows are also significantly stronger compared to those found in more quiescent areas in the same AR. Importantly, the strong turbulent flows are persistent in the loop at early times during the first phase (at 01:54\,UT; Figure\,\ref{fig:cont1}(e)). In Figure\,\ref{fig:cont1}(d) we plot a sample spectrum from this period (excess width of 55\,km\,s$^{-1}$; green curve), which is evidently broader compared to the profile from a quiescent region (excess width of 17\,km\,s$^{-1}$; black curve). At later stages, during the second phase, the emergent spectral profiles show even enhanced line widths at higher intensities (excess width of 82\,km\,s$^{-1}$; red curve).

To quantify and visualize the temporal evolution of turbulent motions with respect to microflare intensity, we color-code the Si\,{\sc iv}\ intensity curve (Figure\,\ref{fig:cont1}(c)), with the excess line widths. This plotting scheme clearly shows that in the early phases, the enhanced turbulent motions are roughly in the range of 40\textendash60\,km\,s$^{-1}$ (see also Figure\,\ref{fig:cont1}(e)). We illustrate the persistence of this enhanced turbulence with individual spectral profiles in Figure\,\ref{fig:spect1} (see also Figure\,\ref{fig:spect2}). These spectral lines are broad and some even exhibit enhancements in blue or red wings, revealing a complex system of turbulent flows in the loop\footnote{The line profiles show predominantly blueshifted asymmetries. Under uniform and equilibrium condition, it is expected that turbulence randomly produce blue and redshifted emission. However, the studied loops and regions are dynamic and continually evolving. These changing conditions coupled with stratification will result in some asymmetries between blue and red wings of the line. The downflows will be halted by the denser material in the lower atmosphere, whereas the upflows may freely expand into low-density upper layers, resulting in a net blueshifted asymmetric line profile.}. 

Clearly, the confined motions in the loop are stronger than the typical non-thermal speeds of 15\textendash20\,km\,s$^{-1}$ observed in ARs \citep[][]{2015ApJ...799L..12D,2015ApJ...810...46H} and are comparable to plasma injection events in loops \citep[][]{2019A&A...626A..98L}. In general, their speeds are comparable to the local sound speed of roughly 50\,km\,s$^{-1}$ at 0.1\,MK in the transition region \citep[][]{1992str..book.....M}. These transonic turbulent plasma motions would carry sufficient energy to heat the loop \citep[][]{1998ApJ...505..957C}, suggesting that the initial heating phase is mostly turbulence-driven. 

During the latter phase, the enhanced turbulent flows are significantly stronger than the local sound speed, reaching Mach numbers of 2 or higher. The broad spectral profiles during this period show plasma emission at Doppler shifts in excess of $\pm200$\,km\,s$^{-1}$ (see Figure\,\ref{fig:cont1}(d)), similar to outflows associated with magnetic reconnection in solar UV bursts \citep[][]{2018SSRv..214..120Y}. The loop brightens near-simultaneously along its length, a feature shared also by much larger loop systems in AR cores \citep[][]{2018A&A...615L...9C}, suggesting the presence of reconnection flows all along the loop. Qualitatively similar results are also obtained for another example of a microflare (see Appendix\,\ref{sec:case2}; Figures\,\ref{fig:cont2},\,\ref{fig:cont3},\,\ref{fig:spect3},\,\ref{fig:spect4}). We find a strong correlation between the intensity and turbulent flows in these microflares (see Appendix\,\ref{sec:ivse}; Figure\,\ref{fig:hist}). Furthermore, excess line widths of Si\,{\sc iv} 1403\,\AA\ and O\,{\sc iv} 1401\,\AA\ line pair are correlated, hinting at a possible multi-thermal nature of turbulence  (see Appendix\,\ref{sec:sio}; Figure\,\ref{fig:sio}).

A gradual spatial broadening of the microflaring loop is observed. In Figure\,\ref{fig:space} we illustrate this in the space-time map of SJI\,2796\,\AA\ intensity along a cut perpendicular to the loop system, closer to its center (where spectroscopic observations are available). Plasma emission from the loop is seen as a clear Gaussian-like enhancement over a non-uniform background. Then to extract the FWHM of the loop, we fit the intensity profile along the cut using a double-Gaussian including a second order polynomial to subtract the non-uniform background and contribution from any adjacent features. The resulting FWHM of the loop segment shows an overall (gradual) increasing trend from 1 to 1.6\,Mm in about 7\,minutes, with superimposed fluctuations (red curve in Figure\,\ref{fig:space}). The increasing trend in loop width (detected in SJI images) correlates with the increasing trend of excess line broadening (obtained using Si\,{\sc iv}\ line profiles, cyan curve) \citep[][]{2008ApJ...686.1372C}. The correlation suggests that the turbulent plasma motions play a role in the broadening of the loop.

To identify the magnetic driver of enhanced turbulence in the loop, we analyzed the HMI magnetic field maps. The loop is rooted in a complex magnetic environment at the solar surface (Figure\,\ref{fig:mag}(a)). The western footpoint is connected to a large patch of negative polarity magnetic field. The eastern footpoint is rooted in a region hosting mixed-polarity magnetic field, which is seen as a 4\,Mm long bipole. The middle section of the loop is apparently rooted in another patch of negative magnetic polarity. Therefore, the eastern half of the loop appears wider compared to the rest of the loop (see Figure\,\ref{fig:cont1}). We observed that the interaction between the two polarities at the eastern footpoint led to magnetic flux cancellation at the surface (Figure\,\ref{fig:mag}(b); similar mixed magnetic polarity fields and flux cancellation events are observed at the footpoints of a microflaring loop discussed in Appendix\,\ref{sec:case2}). The reversal of a component of magnetic field in the process will induce plasma outflow from the reconnection region that drives turbulence through the Kelvin-Helmholtz instability \citep[][]{1999ApJ...517..700L,2009RMxAC..36...81L}, which is demonstrated through numerical simulations \citep[][]{2019arXiv190909179K}.

\section{Turbulence-induced Reconnection}\label{sec:turb}

The enhanced broadening of spectral lines (a factor of 2 to 3 larger than in quiescent regions) is caused by persistent turbulent plasma motions, prior to the onset of microflares (Figure\,\ref{fig:cont1}; see also Figures\,\ref{fig:cont2} and \ref{fig:cont3}). These turbulent flows are externally driven by interacting magnetic elements (e.g., at the eastern loop footpoint as shown in Figure\,\ref{fig:mag}). MHD turbulence induces a weak stochasticity in the magnetic field that triggers fast reconnection \citep[][]{1999ApJ...517..700L}. The turbulent reconnection theory predicts width of current sheets, which we evaluate here. Given a current sheet of length, $L_x$, and energy injection scale of $l_{\mathrm{inj}}$, for the condition $l_{\mathrm{inj}}<L_x$, the width of a current sheet is given by
\begin{equation}
\Delta\approx L_x\frac{U_{\mathrm{obs}}}{v_\mathrm{A}}\left(\frac{l_{\mathrm{inj}}}{L_x}\right)^{\!\!1/2},
\label{eq:wid}
\end{equation}
where $U_{\mathrm{obs}}$ is the observed 1/$e$ width of turbulent motions and $v_\mathrm{A}$ is the Alfv\'{e}n velocity \citep[][]{1999ApJ...517..700L}. Then the FWHM of the current sheet is $2\sqrt{\mathrm{ln}2}\Delta$. 

In the early phases of microflare, the observed turbulent flow speeds are in the range of 40\textendash60\,km\,s$^{-1}$ and the estimated $v_\mathrm{A}\approx1900$\,km\,s$^{-1}$ (details given in Appendix\,\ref{sec:alf}). Based on the observation that the investigated loop brightens all along its length, we assume that the loop length of 14\,Mm to be the length of a current-sheet. 3D MHD simulations of AR loops do show elongated current systems spanning the length of the loop \citep[][]{2017A&A...607A..53W}. The energy injection is on the scale of the size of interacting magnetic bipole (4\,Mm; Figure\,\ref{fig:mag}).\footnote{When observed at even higher spatial resolution, the bipole may reveal magnetic sub-structures unresolved by \textit{SDO}/HMI \citep[][]{2017ApJS..229....4C}. These features could add further complexities to turbulence injection. However, this does not drastically influence the overall observed size of the bipole, as it is not limited by the spatial resolution of \textit{SDO}/HMI.} Then the predicted FWHM of a current sheet is in the range of 250\textendash400\,km. For comparison, for the same values of current-sheet length and Alfv\'{e}n velocity, classical resistivity results in a current-sheet width of roughly 30\,m. Therefore, turbulent reconnection theory predicts a current sheet that is at least three orders of magnitude wider than the classical resistivity case and its value is close to (within a factor 4 to 5) the observed FWHM of the loop segment around 01:54\,UT (see Figure\,\ref{fig:space}). Turbulent broadening of the current sheet can account for a good fraction of the width of the reconnecting loop in the initial stages. It is quite plausible that multiple current sheets, as in models of loops driven by footpoint motions \citep[][]{2008ApJ...677.1348R}, thread the microflare loop system in the initial stages to give rise to the observed loop width. During the abrupt rise phase, $v_\mathrm{A}$ has an upper limit of 750\,km\,s$^{-1}$ (Appendix\,\ref{sec:alf}), when the excess turbulent flow speeds are around 80\,km\,s$^{-1}$. The predicted FWHM of a current sheet is then roughly 1.3\,Mm, which is consistent with the observed FWHM of the loop segment. 

Using HMI data we found that microflare loops are rooted in complex photospheric magnetic settings with interacting mixed magnetic polarities (see Section\,\ref{sec:prop}). These interactions will form a current sheet that ultimately leads to reconnection \citep[][]{2018ApJ...862L..24P,2019ApJ...872...32S}. It is well known from the theory of turbulent flows that there can be different causes that trigger the initial turbulence. The explosive release of magnetic energy can arise from the turbulent reconnection in low-beta plasmas as the injection of turbulence increases the thickness of the outflow region \citep[][]{1999ApJ...517..700L,2009RMxAC..36...81L}. The latter increases the level of turbulence, establishing a positive feedback. This process is consistent with what we observed. Therefore, the estimates of current sheet width based on observationally constrained parameters point to a scenario of fast reconnection onset in a current sheet broadened by turbulence \cite[see][for a discussion on the onset of turbulence in a current sheet]{1974SoPh...36..433P,1977ApJ...216..123H}. Furthermore, the close quantitative agreement between the observed width of the reconnecting loop and the predicted current sheet width makes a compelling observational case for the fast magnetic reconnection induced by turbulence \citep[][]{1999ApJ...517..700L}. In this regard, turbulent plasma motions with speeds comparable to sound speed are not necessary for turbulent magnetic reconnection. What is necessary is the presence of strong turbulent motions, which we do observe.

\section{Conclusion} \label{sec:conc}
 
Previous suggestions of turbulence as a trigger process in solar flares and eruptions, based on lower spatiotemporal resolution observations, remained inconclusive \citep[][]{2008ApJ...686.1372C,2013ApJ...774..122H}. Fragmented current sheets and intermittent jets are investigated in a recent study, suggesting further the turbulent nature of magnetic reconnection \cite[][]{2018ApJ...866...64C}. Through high-cadence \textit{IRIS} observations, we quantitatively demonstrated the development and persistence of high-velocity transonic turbulence, turbulent broadening of the loop, and the subsequent microflare activity in the same feature, thus establishing the direct role of turbulence in triggering fast reconnection. 
 
Our results on the turbulent onset of fast magnetic reconnection support suggestions that turbulence might play a role in the energy transport and cascade in solar flares \citep[][]{2017PhRvL.118o5101K,2018SciA....4.2794J}. While 2D MHD simulations show the dominance of tearing modes \citep[][]{2019PhPl...26i2112H}, recent 3D MHD simulations show that in both externally driven (as in microflares discussed here in which the turbulence is externally driven by the interacting magnetic bipole) and internally developed cases, turbulence would suppress the formation of plasmoids, making turbulent reconnection the main reconnection process \citep[][]{2009ApJ...700...63K,2019arXiv190909179K}. Our results support this view as we observed high-velocity transonic turbulence in the early phases, and in concurrence, there is no clear and obvious evidence for plasmoid-like blobs, at least in the investigated microflares. It is also possible that the spatial resolution of \textit{IRIS} is not sufficient to resolve individual plasmoids. While our observations are consistent with turbulence-induced fast reconnection, higher-resolution observations would be useful to further explore the role of turbulent versus plasmoid reconnection at different stages of the development of various explosive events on the Sun.

\acknowledgments

The authors thank the anonymous referee for insightful comments on the manuscript. L.P.C. thanks Hardi Peter, David Pontin, and Yi-Min Huang for useful discussions and acknowledges previous funding from the European Union's Horizon 2020 research and innovation programme under the Marie Sk\l{}odowska-Curie grant agreement No.\,707837. A.L. acknowledges the support by NASA TCAN AAG1967 grant. \textit{IRIS} is a NASA small explorer mission developed and operated by LMSAL with mission operations executed at NASA Ames Research center and major contributions to downlink communications funded by ESA and the Norwegian Space Centre. \textit{SDO} data are the courtesy of \textit{NASA}/\textit{SDO} and the AIA, EVE, and HMI science teams. CHIANTI is a collaborative project involving the University of Cambridge (UK), the NASA Goddard Space Flight Center (USA), the George Mason University (GMU, USA), and the University of Michigan (USA). This research has made use of NASA's Astrophysics Data System.

\facilities{\textit{IRIS}, \textit{SDO}.}

\appendix

\begin{figure}
\begin{center}
\includegraphics[width=0.45\textwidth]{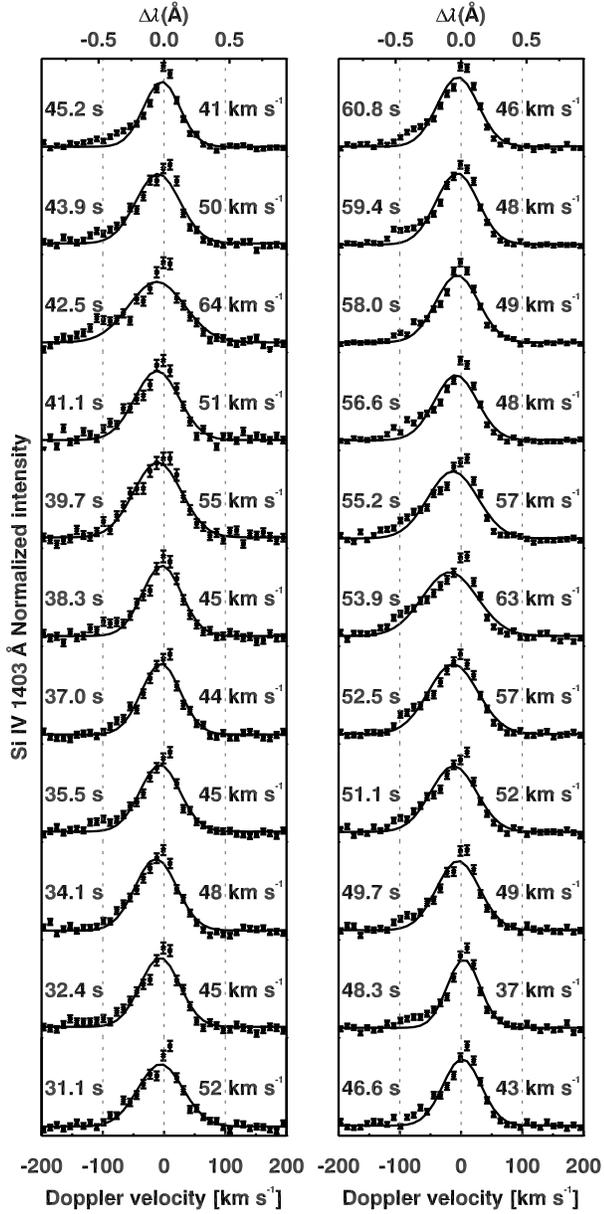}
\caption{Spectral evolution at the site of microflare in AR 12567. This is a continuation of Figure\,\ref{fig:spect1} (main text). The time listed is in seconds elapsed since 01:54\,UT on 2016 July 18.
\label{fig:spect2}}
\end{center}
\end{figure}

\section{Density diagnostics of the microflaring loop using an O\,{\sc iv} line pair}\label{sec:den}

To determine the microflare loop density observed on 2016 July 18, we use an O\,{\sc iv} density-sensitive line pair. We first average the observed line profiles covering a period of 60\,s starting at 01:54\,UT (early phase) and 60\,s starting at 01:56\,UT (abrupt rise phase). The intensity of the O\,{\sc iv} line pair at reference wavelengths of 1401.156 and 1404.812\,\AA\ is then calculated by integrating the respective continuum subtracted average emission, in the Doppler velocity range of $\pm$75\,km\,s$^{-1}$. To determine the electron number density, the observed O\,{\sc iv} 1401 to O\,{\sc iv} 1405\,\AA\ intensity ratio is compared with the theoretical intensity ratios of the same line pair calculated using the CHIANTI atomic database, version\,9 \citep[][]{1997A&AS..125..149D,2019ApJS..241...22D}. 

Based on density diagnostics with the O\,{\sc iv} line pair, the electron number density of the loop at early stages (between 01:54\,UT and 01:55\,UT) is 10$^{17.2}$\,m$^{-3}$, corresponding to a mass density of $2.65\times10^{-10}$\,kg\,m$^{-3}$ (top panel in Figure\,\ref{fig:denmag}). In traditional 1D model atmospheres, such densities are representative of heights of about 2\,Mm above the solar surface \citep[][]{1977ARA&A..15..363W}, suggesting that the loop is a low-lying structure. Such heights are also consistent with models of impulsive flares in which the lower atmosphere is partly heated by accelerated non-thermal electrons \citep[][]{1999ApJ...521..906A}.\footnote{In essence, the height determined from the density diagnostic is primarily used as a reference for obtaining magnetic field strength for Alfv\'{e}n velocity calculation in Appendix\,\ref{sec:alf}. If such densities occur at even higher altitudes in the studied microflares than in traditional 1D model atmospheres, the consequence is an overestimation of Alfv\'{e}n velocity (due to overestimation of magnetic field strength). Such an overestimation of Alfv\'{e}n velocity will in turn result in an underestimation of the predicted current sheet width (Eq.\,\ref{eq:wid} in the main text). Then the actual current sheet width at early stages might be even closer to the observed values of loop FWHM as opposed to a factor of 4 to 5 smaller (see Section\,\ref{sec:turb}).} During the abrupt rise phase (between 01:56\,UT and 01:57\,UT), the ratio of O\,{\sc iv}\ line pair is beyond the theoretical saturation limit on the higher end of densities, and we can only set a lower limit on density diagnostics of the loop of about $10^{18}$\,m$^{-3}$ or mass density of $1.7\times10^{-9}$\,kg\,m$^{-3}$. Such a density enhancement indicates loop heating in the process. In the main text, we suggest that turbulent motions with flow speeds comparable to the local sound speed could heat the plasma in the initial stages.

\section{Determining Alfv\'{e}n Velocity in the Microflaring Loop}\label{sec:alf}

\begin{figure}
\begin{center}
\includegraphics[width=0.48\textwidth]{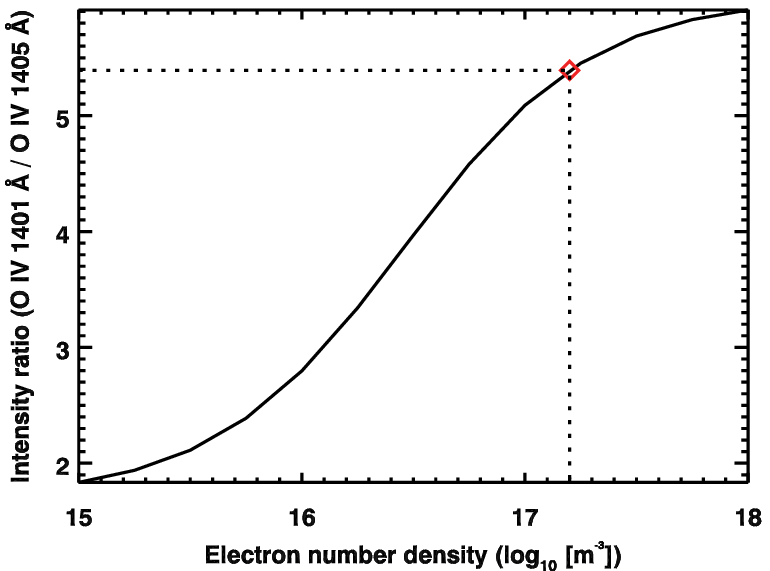}
\includegraphics[width=0.48\textwidth]{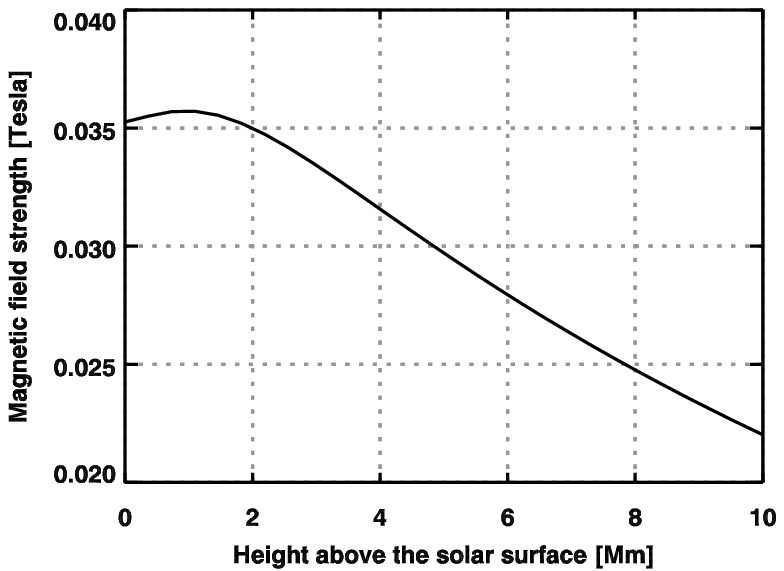}
\caption{Properties of microflaring loop in AR 12567. Top panel: loop density at the early stages of microflaring in AR\,12576 (see Figure\,\ref{fig:cont1}). The solid curve is the electron number density vs. the theoretical intensity ratio of the O\,{\sc iv}\ line pair at 1401 and 1405\,\AA, calculated using the CHIANTI atomic database. The red diamond symbol gives the observed intensity ratio and the corresponding electron number density (see Appendix\,\ref{sec:den} for details). Bottom panel: height variation of magnetic field strength above the solar surface intersecting the microflaring loop. The plot shows horizontal average of magnetic field strength obtained from a potential field extrapolation as a function of height. The spatial averaging is done over the area marked by the white box in Figure\,\ref{fig:mag}. The height 0\,Mm corresponds to the photosphere (see Appendix\,\ref{sec:alf} for details).\label{fig:denmag}}
\end{center}
\end{figure}

To determine Alfv\'{e}n velocity $v_\mathrm{A}=B/\sqrt{\mu_0\rho}$ (where $\mu_0$ is the permeability of vacuum) in the microflaring loop, both plasma density, $\rho$, and magnetic field strength, $B$, are required. For the microflare discussed in the main text, we have density diagnostics from the O\,{\sc iv} line pair (see Appendix\,\ref{sec:den}). To determine magnetic field strength associated with the loop, we used HMI data. We observed that the loop connects a pair of opposite-polarity magnetic field elements at both footpoints; at the eastern footpoint, the loop is rooted in a bipolar region (Figure\,\ref{fig:mag}). To understand how the magnetic field strength varies with height over the loop, we first calculated a 3D potential field by extrapolating the line-of-sight component of the surface magnetic field in the whole AR. In the line of sight intersecting the loop center at a height of 2\,Mm above the surface (nominal loop height), these extrapolations yield a field strength of 0.035\,T (bottom panel of Figure\,\ref{fig:denmag}).

We now have information on the magnetic field strength $B=0.035$\,T and an early-phase plasma mass density $\rho=2.65\times10^{-10}$\,kg\,m$^{3}$ at 2\,Mm height (near the loop center) to calculate the Alfv\'{e}n velocity. Substituting the values for $B$ and $\rho$ yields $v_\mathrm{A}\approx 1900$\,km\,s$^{-1}$. For a 14\,Mm long loop this yields an Alfv\'{e}n crossing time of about 7\,s, which is consistent with fluctuations of duration 5 to 10\,s, seen in the excess line width (see Figure\,\ref{fig:cont1}(e); between 01:54\,UT and 01:56\,UT). Such short-period quasi-periodic non-thermal broadening is also observed at the footpoints of a flaring loop \citep[][]{2018SciA....4.2794J}. Using sound speed $v_\mathrm{S}\approx 50$\,km\,s$^{-1}$ at 0.1\,MK, we find that the plasma$-\beta$ ($=2v_\mathrm{S}^2/\gamma v_\mathrm{A}^2$, where $\gamma$ is the adiabatic index) is of the order of $10^{-3}$ (assuming $\gamma= 5/3$). During the rise phase, a lower limit on plasma density is $\rho\approx\ 1.7\times10^{-9}$\,kg\,m$^{-3}$. This yields an upper limit on Alfv\'{e}n velocity, $v_\mathrm{A}\approx 750$\,km\,s$^{-1}$ and lower limit on plasma$-\beta\approx 5\times10^{-3}$.

\section{Observational Details of a Second Microflare}\label{sec:case2}

\begin{figure*}
\begin{center}
\includegraphics[width=0.75\textwidth]{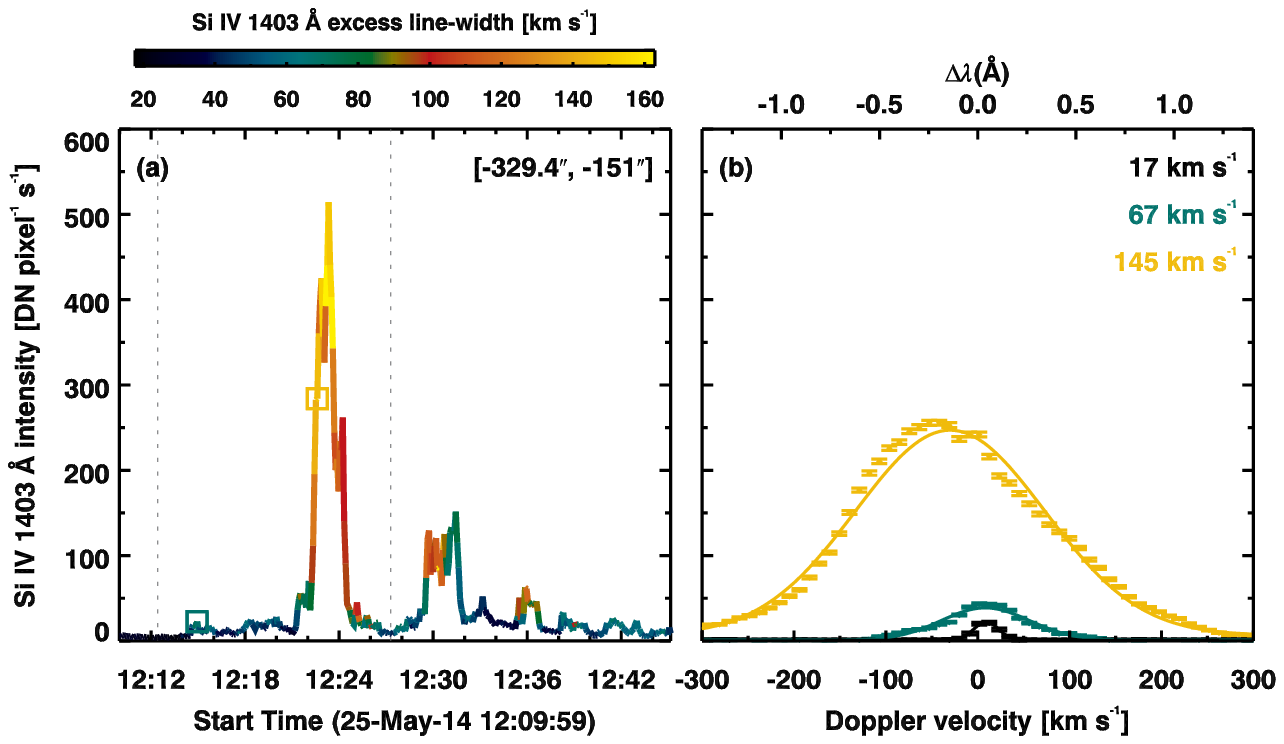}
\includegraphics[width=0.75\textwidth]{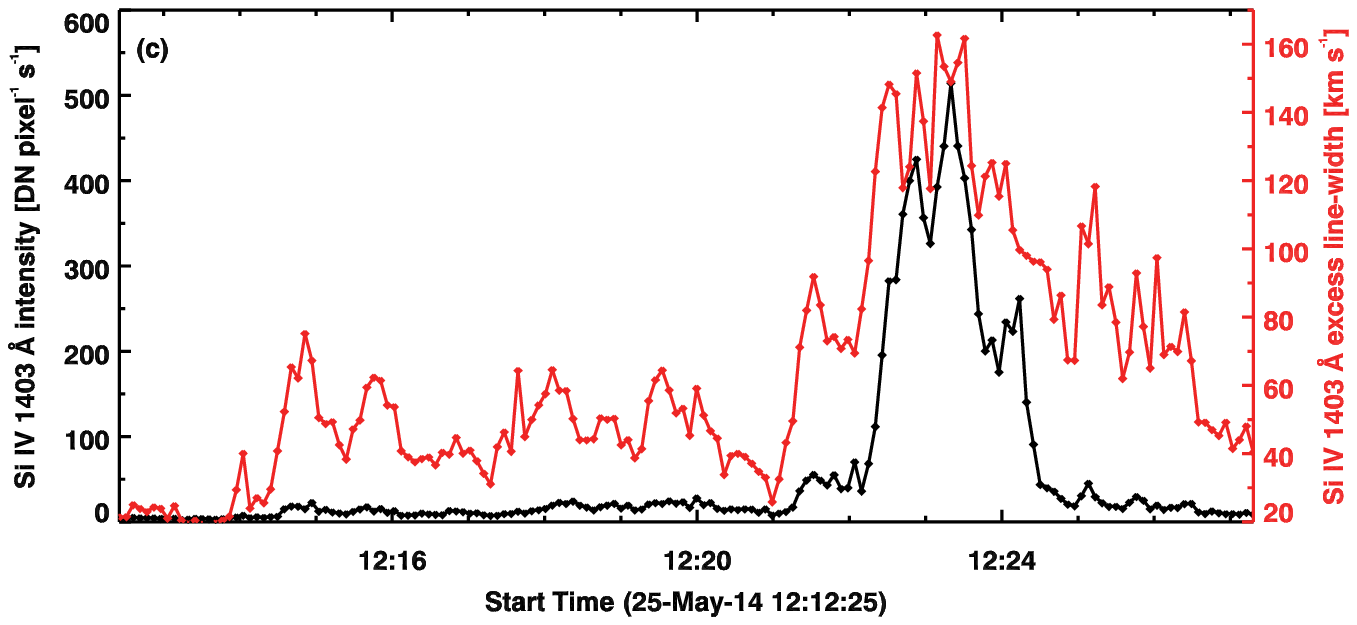}
\caption{Repeated microflare in the core of AR\,12073. Panels (a)\textendash(c) are same as panels (c)\textendash(e) in Figure\,\ref{fig:cont1} (main text), but plotted for a microflaring loop observed in AR\,12073 on 2014 May 25. The spatial location of these diagnostics is marked by a red square on a context image of the loop displayed in the top-right panel of Figure\,\ref{fig:hist}. In panel (b), the solid black curve is the observed line profile obtained by spatially averaging the Si\,{\sc iv} line profiles from a quiet region (multiplied by a factor of 3) marked with vertical green bars in the top-right panel of Figure\,\ref{fig:hist}. The symbols are observations with only every second point in the spectrum plotted along with the photon noise (see Section\,\ref{sec:obs} and Appendix\,\ref{sec:case2} for details).\label{fig:cont2}}
\end{center}
\end{figure*}

\begin{figure*}
\begin{center}
\includegraphics[width=0.75\textwidth]{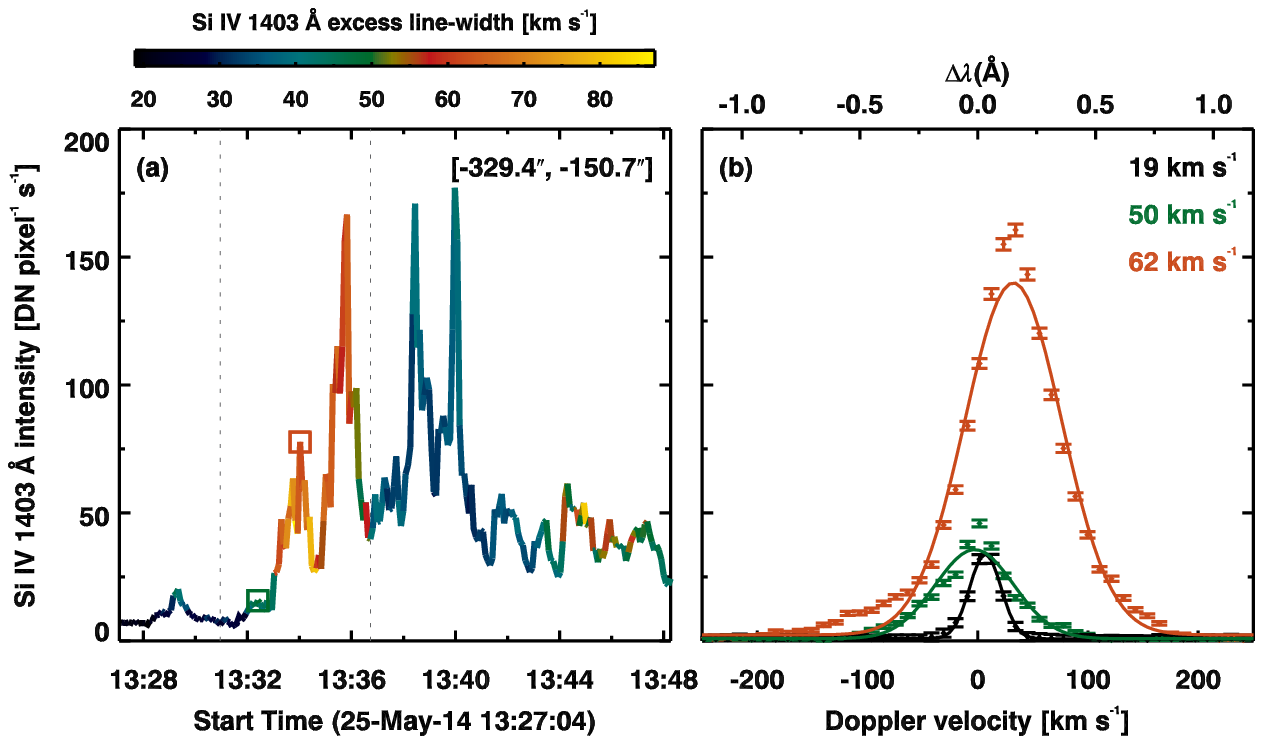}
\includegraphics[width=0.75\textwidth]{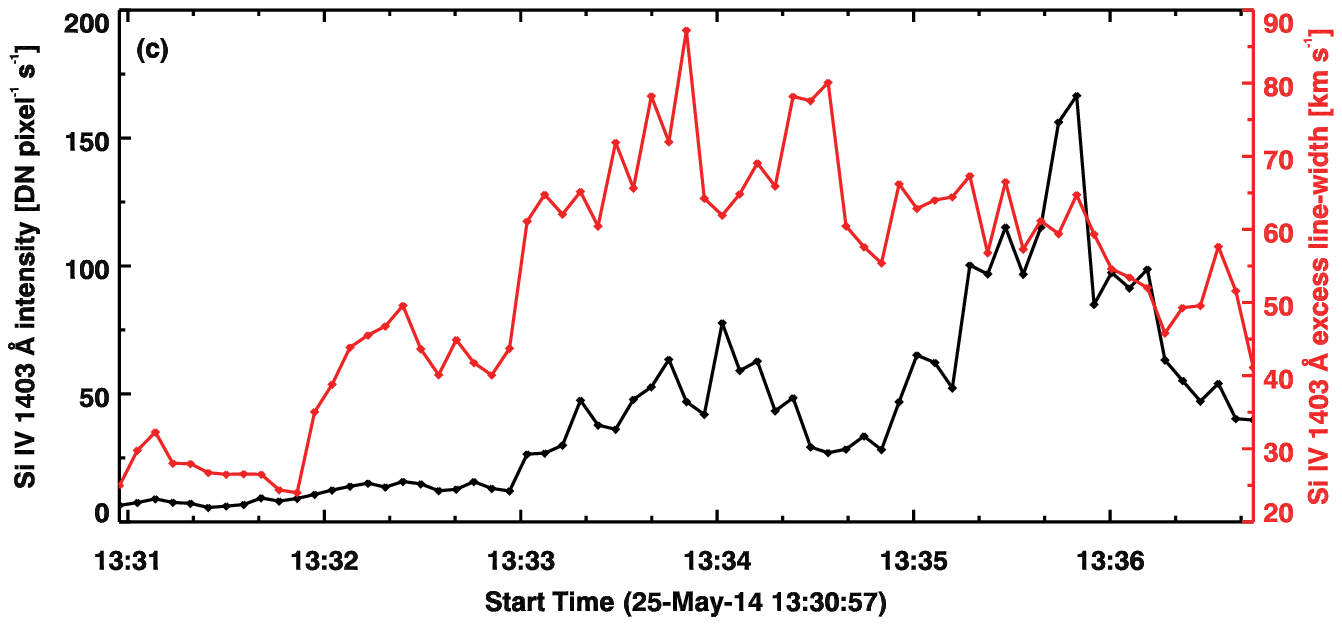}
\caption{Another example of a repeated microflare in the core of AR\,12073. Same as Figure\,\ref{fig:cont2} but plotted for a different repeated microflare (see Section\,\ref{sec:obs} and Appendix\,\ref{sec:case2} for details).\label{fig:cont3}}
\end{center}
\end{figure*}

\begin{figure}[t]
\begin{center}
\includegraphics[width=0.45\textwidth]{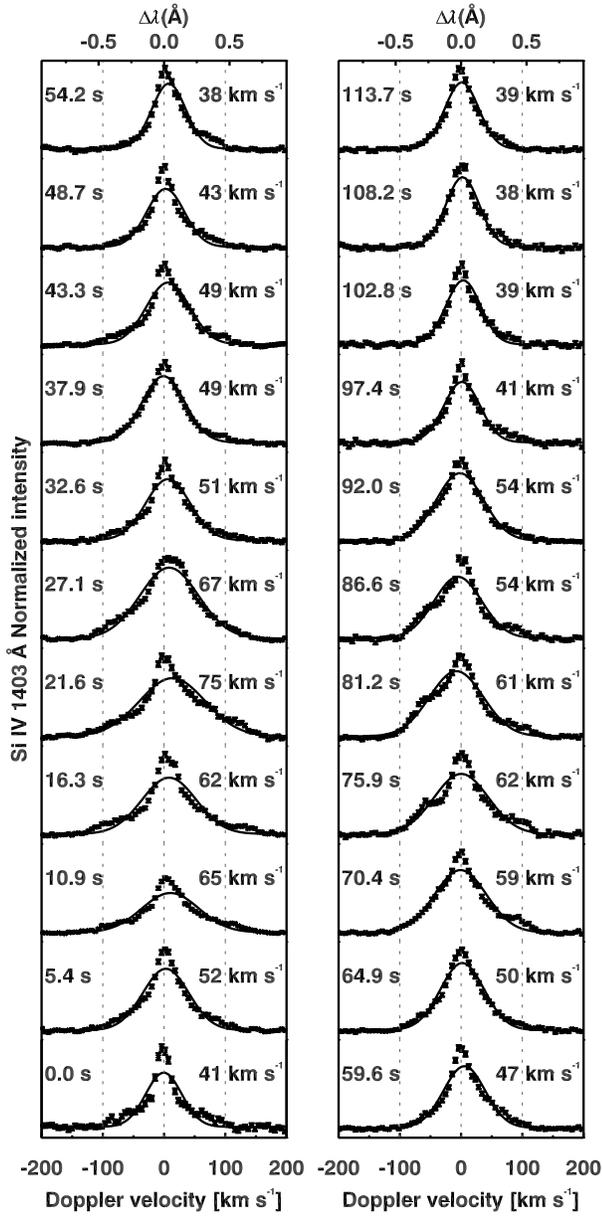}
\caption{Same as Figure\,\ref{fig:spect1} but plotted for the microflare event observed in AR\,12073 on 2014 May 25, starting at 12:14:29\,UT.
\label{fig:spect3}}
\end{center}
\end{figure}

\begin{figure}[t]
\begin{center}
\includegraphics[width=0.45\textwidth]{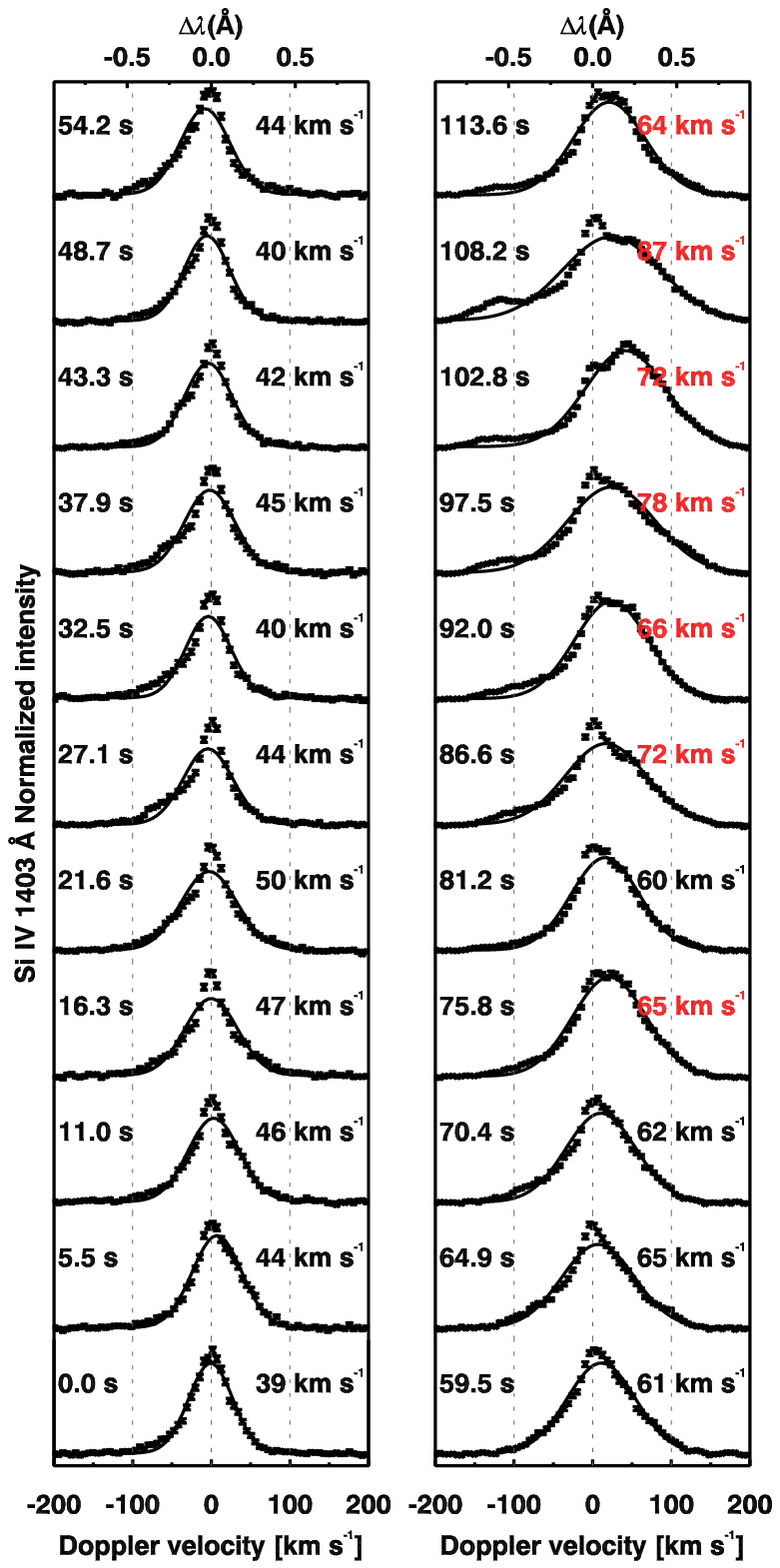}
\caption{Same as Figure\,\ref{fig:spect1} but plotted for the microflare event observed in AR\,12073 on 2014 May 25, starting at 13:32:02\,UT. The profiles in the right column cover the rapid rise phase of the microflare.\label{fig:spect4}}
\end{center}
\end{figure}

In the main text, we discussed the results obtained from a non-repeating microflare. Here we discuss a second microflare. Unlike the first case, the microflare discussed here exhibited repeated brightenings. On 2014 May 25 between 11:51\,UT and 14:21\,UT, \textit{IRIS} observed AR\,12073, located away from disk center at $(-330\arcsec, -151\arcsec)$, in sit-and-stare mode at a roll angle of 5$^{\circ}$. The spatial position, $\mu$, of the AR is 0.925. The viewing angle, $\theta$, between the observer's line of sight
and the normal to the local solar surface of the AR is $\theta = \rm{cos}^{-1}(\mu) \approx 22^{\circ}$. These pointing details are as per the \textit{IRIS} SJI header information at the start of observations. The reference coordinates of the \textit{IRIS} spectrographic slit are $(-329.5\arcsec, -151.9\arcsec)$. The \textit{IRIS} slit captured a repeated microflaring loop of length 16\,Mm in the core of the AR, with each event lasting for a few minutes. The microflare is imaged by the 1400\,\AA\ passband of \textit{IRIS} SJI (sampling temperatures of 5000\,K to 0.1\,MK from the upper photosphere to the transition region). These SJI images have a cadence of 11\,s. The spectroscopic data have a cadence of 5.4\,s, with an exposure time of 4\,s, and spatial scale of 0.166\arcsec\,pixel$^{-1}$ along the slit. These data are re-binned to 0.33\arcsec\,pixel$^{-1}$ along the slit. The spectral sampling of the Si\,{\sc iv}\ line at 1403 Å is 25 m\AA. In this case, the continuum intensity to be subtracted is obtained by averaging emission from two wavelength windows, one between 1400.17 and 1400.82\,\AA\  and the second between 1405.21 and 1405.51\,\AA. The spectroscopic properties of this repeated microflare are plotted in Figures\,\ref{fig:cont2}, \ref{fig:cont3}, \ref{fig:spect3}, and \ref{fig:spect4}.

\section{Further Plasma Characteristics of Turbulent Reconnection}
\subsection{Intensity versus Excess Line Width}\label{sec:ivse}

\begin{figure*}
\begin{center}
\includegraphics[width=0.8\textwidth]{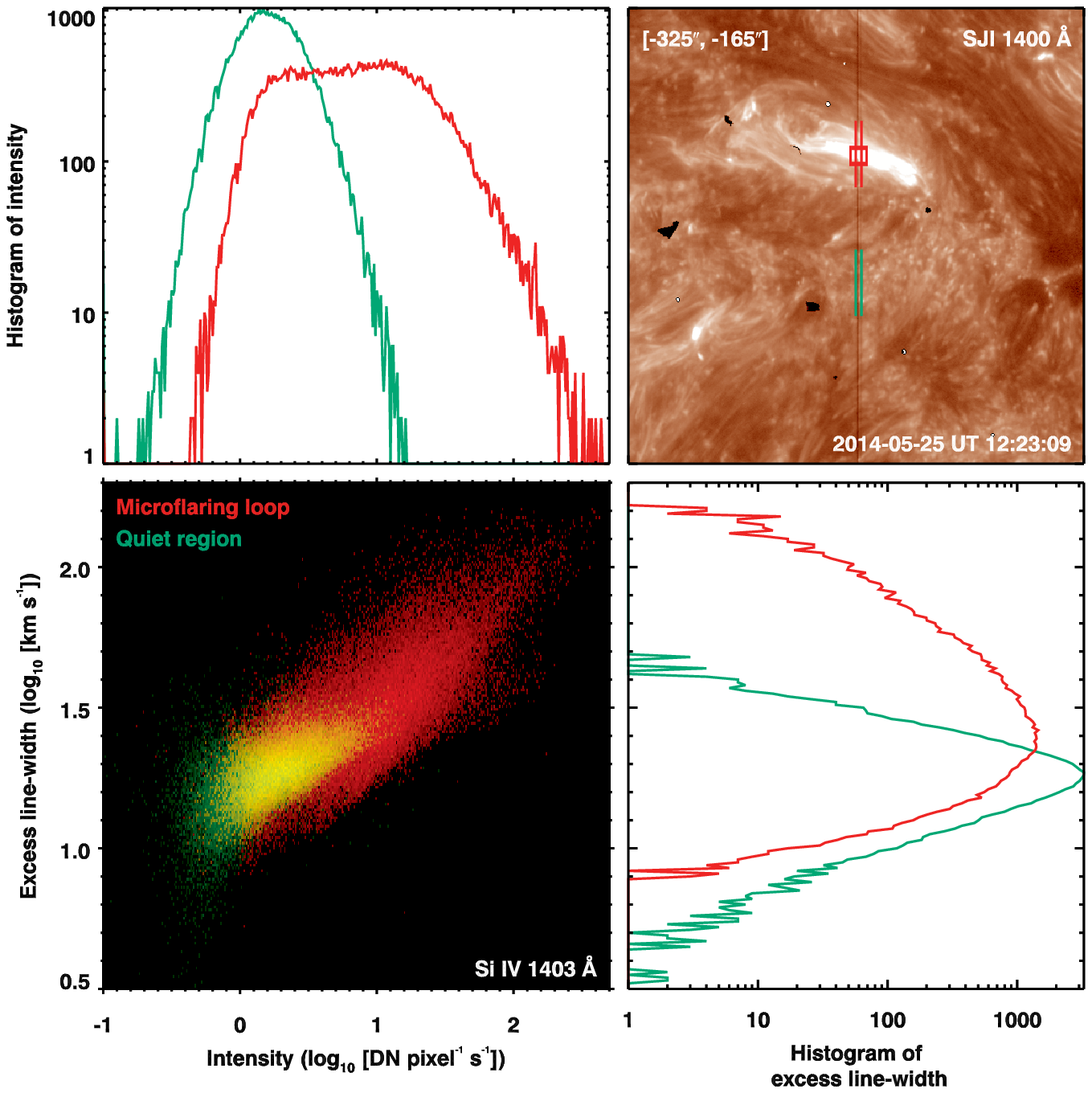}
\caption{Plasma properties of a microflaring loop in the transition region to solar corona at 0.08\,MK. The top-right panel shows a snapshot of AR\,12703 observed on 2014 May 25 at 12:23\,UT, from the \textit{IRIS} SJI 1400\,\AA\ passband. The field of view of $80\arcsec\times80\arcsec$ is centered at ($-325\arcsec, -165\arcsec$) on the Sun. The red and green colored bars mark the locations of a microflaring loop and a quiet region in the AR core. The spectroscopic properties of the loop plotted in Figures\,\ref{fig:cont2}, \ref{fig:cont3}, \ref{fig:spect3} and \ref{fig:spect4} are from the center of the red-squared regions within the loop. The lower-left panel is a 2D PDF of the Si\,{\sc iv} 1403\,\AA\ line intensity vs. its excess line width. The PDF in green is for the quiet region and the red-colored PDF is for the microflaring loop. These PDFs cover a period of 150\,minutes from 11:51\,UT to 14:21\,UT on 2014 May 25. The individual histograms of the intensity and excess line width are shown in the top-left and bottom-right panels, respectively (see Appendices\,\ref{sec:case2} and \ref{sec:ivse} for details).
\label{fig:hist}}
\end{center}
\end{figure*}

We generalize the relation between the line intensity and excess line broadening and find a strong correlation between the two line parameters in the microflaring loop (Figure\,\ref{fig:hist}). In quiescent part of the AR (green points and curves), our analysis reproduces the well-known correlation between intensity and non-thermal broadening observed both in quiet Sun and ARs at transition region temperatures of 0.1\,MK \citep[][]{1998ApJ...505..957C,2015ApJ...799L..12D}. In the quiescent region, the distribution of excess line widths peak at about 20\,km\,s$^{-1}$, and at these values, the 2D probability density function (PDF) of intensity versus excess line width saturates. On the other hand, the microflaring loop displays a similar correlation that extends to larger values both in intensity and excess line widths. At lower intensities (say, below 50\,DN\,pixel$^{-1}$\,s$^{-1}$), there is an overlap between the quiescent region and microflaring loop intensity versus excess line width PDFs. Nevertheless, the loop displays a larger scatter skewed toward higher values of turbulent motions (in excess of 50\,km\,s$^{-1}$). In brightening loops, such low intensities are typically observed at the early phases of microflaring activity (see Figures\,\ref{fig:cont1}, \ref{fig:cont2}, and \ref{fig:cont3}). The PDFs presented in Figure\,\ref{fig:hist} provide further evidence for the presence of enhanced turbulence at the onset of reconnection.

\subsection{Spectral Comparison of Excess Line Width}\label{sec:sio}
\begin{figure}
\begin{center}
\includegraphics[width=0.4\textwidth]{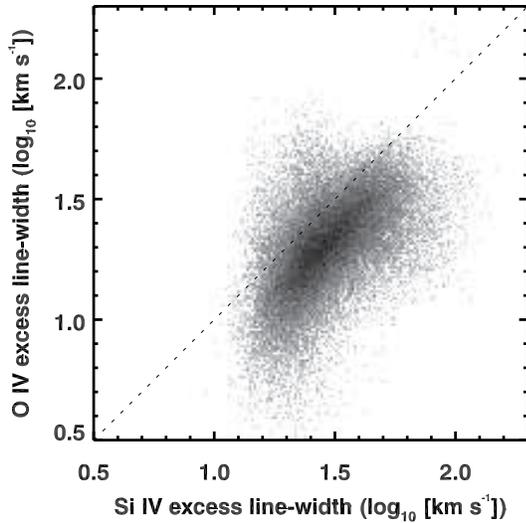}
\caption{Spectral comparison of excess line width. The plot shows the 2D PDF of excess line widths of Si\,{\sc iv} 1403\,\AA\ vs.\,O\,{\sc iv} 1401\,\AA\ line pair for the microflaring loop in AR\,12703. The spatial and temporal sampling are the same as in the case of red-colored PDF in Figure\,\ref{fig:hist} (see Appendices\,\ref{sec:case2} and \ref{sec:sio} for details).
\label{fig:sio}}
\end{center}
\end{figure}

Using the combination of Si\,{\sc iv} and O\,{\sc iv} lines, we can probe multi-thermal characteristics of plasma turbulence between 0.08 and 0.14\,MK (i.e. between the equilibrium temperatures of respective atomic species). To this end, we consider the example of microflaring loop in AR\,12703 and fit the Si\,{\sc iv} 1403\,\AA\ and the O\,{\sc iv} 1401\,\AA\ line pair with two single Gaussians, simultaneously, to derive characteristics of turbulent flows. The resulting excess line widths are plotted as a 2D PDF in Figure\,\ref{fig:sio}. There is a correlation in the excess line widths of the line pair. This suggests a multi-thermal turbulent flow system in the plasma. 

The excess line widths of the O\,{\sc iv} 1401\,\AA\ line, however, are systematically lower than those of the Si\,{\sc iv} 1403\,\AA\ line. This could mean a lower strength of turbulent flows at 0.14\,MK\ compared to those at 0.08\,MK. However, this is unlikely because the trend is inverse to the observations of quiet Sun that show systematic increase of excess line widths from 0.01 to 0.3\,MK \citep[][]{1998ApJ...505..957C}. Instead, the lower values of the excess line widths of the O\,{\sc iv} 1401\,\AA\ line could be due more likely to weakening of that line as a result of collisional de-excitation at higher densities, compared to the Si\,{\sc iv} line, under flaring conditions. Moreover, for the observed range of electron number densities between $10^{17}$ and $10^{18}$\,m$^{-3}$ (see Appendix\,\ref{sec:den}), the radiative power of the O\,{\sc iv} 1401\,\AA\ line itself reduces by a factor of 5 \citep[][]{1977ApJ...215..652F}. This combination of weak intensity in comparison with Si\,{\sc iv} line and further reduction of radiative power with increasing density could result in lower excess line widths of the O\,{\sc iv} 1401\,\AA\ line. Though the observations hint at multi-thermal turbulence in the microflaring loops, the O\,{\sc iv} line alone may not be a suitable diagnostic to investigating the turbulent nature of plasma at the reconnection onset.

\end{document}